\documentclass[%
reprint,
superscriptaddress,
amsmath,amssymb,
aps,
pre
]{revtex4-2}

\usepackage{mathrsfs}
\usepackage{dsfont}
\usepackage{xcolor}
\usepackage{fontawesome}
\usepackage{graphicx}
\usepackage{dcolumn}
\usepackage{bm}
\usepackage{hyperref}
\hypersetup{breaklinks=true}

\definecolor{red}{rgb}{0.75,0,0}
\definecolor{blue}{rgb}{0,0,0.75}
\definecolor{green}{rgb}{0,0.5,0}

\newcommand{\di}[1]{\mathrm{di}\left(#1\right)}

\usepackage{blindtext}

\begin{document}
	\title{Localized spatiotemporal dynamics in active fluids}
	
	\author{Luca Barberi}%
	\thanks{Current address: Medicines for Malaria Venture, Geneva, Switzerland}
	\email{luca.barberi@unige.ch}
	\affiliation{Department of Biochemistry, University of Geneva, 1211 Geneva, Switzerland}
	\affiliation{Department of Theoretical Physics, University of Geneva, 1211 Geneva, Switzerland}
	\author{Karsten Kruse}%
	\email{karsten.kruse@unige.ch}
	\affiliation{Department of Biochemistry, University of Geneva, 1211 Geneva, Switzerland}
	\affiliation{Department of Theoretical Physics, University of Geneva, 1211 Geneva, Switzerland}
	
	\date{\today}
	
	\begin{abstract}
	From cytoskeletal networks to tissues, many biological systems behave as active materials. Their composition and stress-generation is affected by chemical reaction networks. In such systems, the coupling between mechanics and chemistry enables self-organization, for example, into waves. Recently, contractile mechanochemical systems were shown to be able to spontaneously develop localized spatial patterns. Here, we show that these localized patterns can present intrinsic spatiotemporal dynamics, including oscillations and chaotic-like dynamics. We discuss their physical origin and bifurcation structure.
	\end{abstract}

	\maketitle
	
\section{Introduction}
Self-organized oscillations are a common feature of living systems spanning a large range of scales, from genetic networks to animal populations~\cite{murray1993}. In cells and tissues, such oscillations correlate with vital processes such as migration and cell division~\cite{beta2017}. Although commonly analyzed in a pure chemical context, cellular oscillations often involve the mechanical units of the cytoskeleton~\cite{beta2017,wu2021}, a polymeric network consisting notably of actin filaments. From a physical point of view, the actin cytoskeleton is an active fluid, as various cortical processes driven by ATP hydrolysis can generate mechanical stress.
	
Many studies on mechanochemical oscillatory cellular dynamics and closely related waves focus on the actin cortex~\cite{yang2018}. It is a thin layer of actin cytoskeleton beneath the plasma membrane of animal cells. Active cortical contractions coupled to actin-filament turnover can lead to sustained oscillations~\cite{bornens1989,pletjushkina2001,paluch2005,sedzinski2011,solon2009,martin2009,westendorf2013}. Furthermore, the cortex can exhibit spontaneous waves, in particular, in the context of cell migration~\cite{vicker2000,gerisch2004,Weiner2007,stankevicins2020}, but also during cell division~\cite{bement2015}. 

Although cortical oscillations can be of purely mechanical origin~\cite{paluch2005,zumdieck2005,sedzinski2011,dierkes2014}, the coupling of cytoskeletal dynamics to biochemical signaling networks is typically considered to be essential. Indeed, molecular networks establishing mutual feedback between the activity of signaling proteins and filament assembly as well as cytoskeletal contraction have been identified~\cite{bement2015,graessl2017,michaud2022} and turn the actin cortex into an excitable medium~\cite{michaux2018,iglesias2012,ecker2021}. 

Even in absence of active stress generation, the coupling between signaling modules and the cytoskeleton can lead to spontaneous (polymerization) waves~\cite{doubrovinski2008,whitelam2009,carlsson2010,bernitt2017,yochelis2022}. Still, the fascinating dynamics of the actin cortex has led to the introduction and study of several theoretical descriptions combining the physics of active fluids and the dynamics of activity regulators~\cite{bois2011,radszuweit2013,kumar2014,banerjee2017,nishikawa2017, bhattacharyya2021, staddon2022,deljunco2022,barberi2023}.   In addition to cortical waves and oscillations that span the whole cell surface,  experiments in adherent cells have identified oscillations that are localized in space~\cite{baird2017,graessl2017}. The mechanism underlying these dynamic states is currently unknown. 

Motivated by the  dynamics of the actin cortex, we study here a hydrodynamic description of an active fluid coupled to a reaction network that controls fluid turnover.  We have recently shown that this framework can lead to localized stationary states (LSSs)~\cite{barberi2023}. Here, we extend our previous results and show that localized oscillatory states (LOSs) can spontaneously emerge in such mechanochemical systems. In Sec.~\ref{sec:model}, we introduce our description. We report different types of LOSs in one spatial dimension in Sec.~\ref{sec:oneD} and discuss the underlying mechansims. In Sec.~\ref{sec:twoD}, we explore LOSs in two spatial dimensions and, in particular, find chiral LOSs. We conclude by discussing the generality of our results and how they might help to understand cortical actin structures.
	
\section{Mechanochemical theory of the cell cortex}\label{sec:model}

To describe the dynamics of an active fluid coupled to a biochemical network, we combine active hydrodynamics with a reaction-diffusion system~\cite{deljunco2022,barberi2023}. The biochemical network is regulating assembly of the active material. In the case of the cortical actomyosin network in animal cells, this network would include small GTPases from the Ras and Rho families. These proteins exist in active and inactive states, where the former are typically attached to the cellular membrane and the latter are dissolved in the cytoplasm. In their active form, these small GTPases regulate factors promoting the nucleation and growth of actin filaments like formins or the Arp2/3 complex. In turn, the actomyosin network feeds back on the small GTPases' activity. 

Since here we are  interested in studying generic properties of mechanochemical systems, we refrain from trying to give a comprehensive description of the biochemical regulatory network and their coupling to the cytoskeleton. Instead, we consider the active and inactive forms of some nucleation promoting factor -- nucleator for short -- and the active fluid. For the actin cortex it is appropriate to use an effective one-component description~\cite{joanny2013}. We thus introduce the corresponding densities $N_a$, $N_i$, and $C$. The time evolution of these densities is governed by mass-conservation laws. Explicitly, we use the equations introduced in Ref.~\cite{barberi2023},	
	\begin{align}
		\partial_T C + \nabla \cdot \bm{J}_c &= A N_a - K_d C, \label{eq:actin_dynamics} \\
		\partial_T N_a + \nabla \cdot \bm{J}_a &= \Omega_0 (1 + \Omega N_a^2)N_i - \Omega_d C N_a,  \label{eq:nucleator_dynamics_a}\\
		\partial_T N_i + \nabla \cdot \bm{J}_i &= -\Omega_0 (1 + \Omega N_a^2)N_i + \Omega_d C N_a, \label{eq:nucleator_dynamics_b}
	\end{align}
where all quantities are dimensionless.  As suggested by electron micrographs of the cortex~\cite{fritzsche2016}, we consider the active fluid to be isotropic.

The current $\bm{J}_c = \bm{V} C - \mathscr{D}_c \nabla C$ for the active fluid consists of an advective and a diffusive term. Here, $\mathscr{D}_c$ is a diffusion constant, and $\bm{V}$ denotes the active fluid's velocity field. The currents $\bm{J}_a$ and $\bm{J}_i$ for the active and inactive nucleators have the same form, but with different diffusion constants $\mathscr{D}_a$ and $\mathscr{D}_i$. Since active nucleators are bound to the membrane, whereas their inactive forms reside in the cytoplasm, we will consider $\mathscr{D}_a < \mathscr{D}_i$. In absence of advection, Eqs.~\eqref{eq:actin_dynamics}--\eqref{eq:nucleator_dynamics_b} reduce to a reaction-diffusion system that has been  used to describe actin polymerization waves~\cite{ecker2021}. 

The source and sink terms on the right hand sides account for transitions of nucleators between their active and inactive states as well as for the assembly and disassembly of the active fluid. The signs are chosen such that all corresponding parameters are positive. Specifically, $A$ is the rate of nucleation by active nucleators and $K_d$ the rate of spontaneous active fluid disassembly. The parameter $\Omega_0$ gives the rate of spontaneous nucleator activation. Experiments suggest that there is cooperative nucleator activation~\cite{kamps2020,michaud2022} which we capture by the parameter $\Omega$. Furthermore,  experiments show that there is negative feedback from the actomyosin on nucleator activity~\cite{kamps2020,michaud2022}. We account for this effect through the parameter $\Omega_d$. Note that Eq.~\eqref{eq:nucleator_dynamics_a} and Eq.~\eqref{eq:nucleator_dynamics_b} conserve the total number of nucleators. Consequently, the average total nucleator density, $\bar{N} = (1/\mathcal{V})\int_\mathcal{V} (N_a + N_i) d\mathcal{V}$, where $\mathcal{V}$ is the system volume, is constant. From now on, we consider the  densities $C$, $N_a$, and $N_i$ to be scaled by $\bar{N}$, but keep the same notation as before. This rescaling implies that $\Omega_d\propto \bar{N}$ and $\Omega \propto \bar{N}^2$.  

To close the dynamic equations, we use force balance, which captures momentum conservation in the case of low Reynolds number flows relevant for cortical actin dynamics. We choose~\cite{barberi2023}
	\begin{align}
		\nabla \cdot \Sigma &= \bm{V}, \label{eq:force_balance}\\
		\Sigma &= 2 \mathsf{V} +\left[\nabla\cdot\bm{V} + \Pi(C)\right]\mathds{1}, \label{eq:stress_field}
	\end{align}
where $\Sigma$ is the stress tensor, $\mathsf{V} = (1/2)[\nabla\bm{V} + (\nabla\bm{V})^T-(\nabla\cdot\bm{V}/d)\mathds{1}]$ is the traceless strain rate tensor, $d$ the number of spatial dimensions and $\mathds{1}$ the identity. Finally, $\Pi(C) = \left(Z C^2 - B C^3\right)\mathds{1}$ accounts for the non-viscous stress. If $Z, B>0$, then the first term can be interpreted as an active contractile stress, which dominates at low fluid densities, whereas hydrostatic contributions captured by the second term dominate at high densities~\cite{joanny2013}. Alternative forms of the non-viscous stress can be used~\cite{bois2011}. Note that the scaling of the densities by $\bar{N}$ implies $Z \propto \bar{N}^2$ and $B \propto \bar{N}^3$.
	
Henceforth, we assume periodic boundary conditions. Our results are qualitatively similar if no-flux boundary conditions are assumed. Numerical solutions of Eqs.~\eqref{eq:actin_dynamics}--\eqref{eq:stress_field} in 1D and 2D are obtained using a custom code written in Julia~\cite{bezanson2017}, available online~\cite{luca_barberi_2023_10209642}. The code uses pseudo-spectral methods to compute spatial derivatives, on a grid of $512$ nodes in 1D and of $512\times512$ nodes in 2D. Time integration is performed using a fixed timestep $\Delta t = 10^{-4}$ and explicit methods: midpoint method for 1D simulations, $O(\Delta t^2)$, and Euler method for 2D simulations, $O(\Delta t)$. Unless specified otherwise, we use the parameter values: $\mathscr{D}_c = 0.01$, $\mathscr{D}_a = 0.1$, $\mathscr{D}_i = A = K_d = 1$, $\Omega_d/\bar{N} = 10$, $B / \bar{N}^3 = Z / \bar{N}^2$, $\bar{N} = 1$. In Sect.~\ref{sec:oneD}, we consider the dynamic Eqs.~\eqref{eq:actin_dynamics}--\eqref{eq:stress_field} in one spatial dimension. In Sect.~\ref{sec:twoD}, we treat the case of two spatial dimensions.
	
\section{Localized spatiotemporal dynamics in one dimension}\label{sec:oneD}

	\begin{figure}
		\includegraphics[width=0.75\columnwidth]{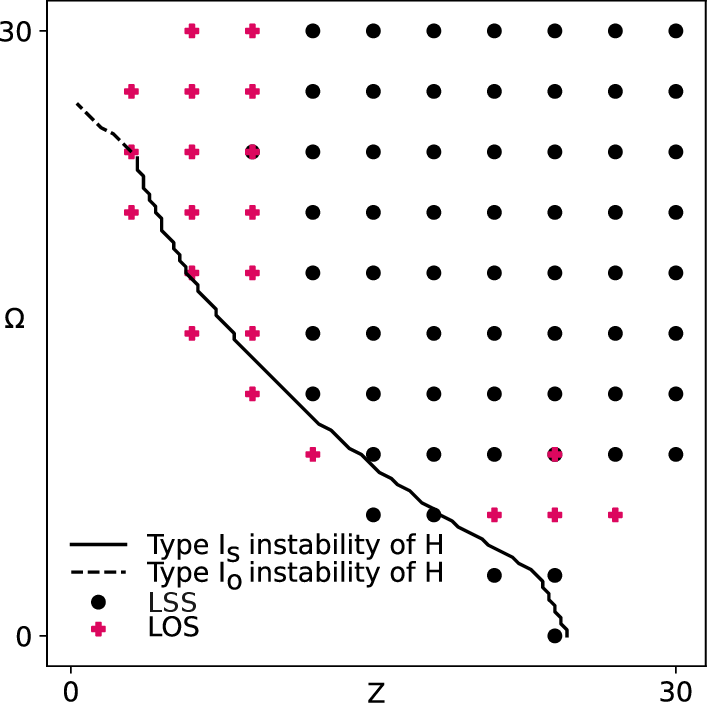}
		\caption{\label{fig:fig_1} Phase diagram of localized states in one dimension. The symbols indicate parameter values for which localized states (black circles: stationary, magenta crosses: oscillatory) are numerically obtained through time evolution of a localized initial condition. Stationary and oscillatory localized states can coexist. The lines are reproduced from Ref.~\cite{barberi2023} and limit the region of linear stability of the homogenous stationary state (solid: Type $\mathrm{I_s}$ instability, dashed: Type $\mathrm{I_o}$ instability). The homogeneous state is stable below and unstable above the lines, respectively. Parameter values are $\mathscr{D}_c = 0.01$, $\mathscr{D}_a = 0.1$, $\mathscr{D}_i = A = K_d = 1$, $\Omega_d/\bar{N} = 10$, $B / \bar{N}^3 = Z / \bar{N}^2$, $\bar{N} = 1$, $L = 10\pi$. Numerical solutions are obtained on a grid with $512$ sites.}
	\end{figure}
Consider a system of length $L = 10\pi$. Equations~\eqref{eq:actin_dynamics}--\eqref{eq:stress_field} admit a unique homogenous state H with $C=AN_a/K_d$, $V=0$, $N_i = 1 - N_a$ and $N_a$ such that the right hand side of Eq.~\eqref{eq:nucleator_dynamics_a} vanishes~\cite{ecker2021}. A numerical linear stability analysis, performed in Ref.~\cite{barberi2023}, shows that H becomes unstable if $Z$ and $\Omega$ exceed critical values, Fig.~\ref{fig:fig_1}. The instability occurs at a finite wavelength, and can be either stationary  or oscillatory, respectively, Type $\mathrm{I_s}$ and Type $\mathrm{I_o}$ in the language of Ref.~\cite{cross1993}. Specifically, the system has a Type $\mathrm{I_s}$ instability if it is unstable against infinitesimal perturbations with wavenumber $k \neq 0$ and temporal frequency $2\pi/\tau=0$, where $\tau$ is the temporal period. Type $\mathrm{I_o}$ instabilities are similar, but with $2\pi/\tau \neq 0$.

The Type $\mathrm{I_o}$ instability is driven by the nonlinear chemical network and occurs in the range $0 \leq Z \lesssim 3$. For $Z=0$ the system exhibits excitation waves~\cite{ecker2021}. For higher contractility, $Z \gtrsim 3$, H undergoes a Type $\mathrm{I_s}$ instability. Here, the instability is driven by the contractility and does not require a chemical network. In absence of a chemical network, local maxima in the fluid density emerge that fuse with time, such that eventually all the fluid is concentrated in a small region of space~\cite{bois2011}. In the presence of both, contractility and a nonlinear chemical network, the critical values of the parameters decrease, and the instabilities are mechanochemical. 

The traveling waves as well as the contracted states are spatially extended, in that they affect the fluid and nucleator densities throughout the whole system. In addition, there are localized states~\cite{barberi2023}. The existence of these states requires both, the contraction of the active fluid and the regulation of the fluid density through the chemical nucleator network.

Localized states can be stationary (localized stationary states, LSSs), as discussed in detail in Ref.~\cite{barberi2023}, but also oscillatory (localized oscillatory states, LOSs). They exist in a large region of parameter space, Fig.~\ref{fig:fig_1}. This region mostly overlaps with the region of parameter space where the homogeneous state is unstable. This is a consequence of the slanted snaking instability that generates localized states in our system, as shown in Ref.~\cite{barberi2023}.

LSSs and LOSs can coexist. We find LOSs mostly for low contractility, in the proximity of the Type $\mathrm{I_o}$ instability of H. In the following, we discuss the emergence of LOSs in detail. We identify two distinct mechanism leading to LOSs. Either they emerge through a local instability of H or through a secondary instability of an LSS. The former dominates for low, whereas the latter dominates for high contractility. We discuss both instabilities in turn.

\subsection{Low-contractility localized oscillations}\label{sec:low-contractility_los}
		
\begin{figure}
		\includegraphics[width=\columnwidth]{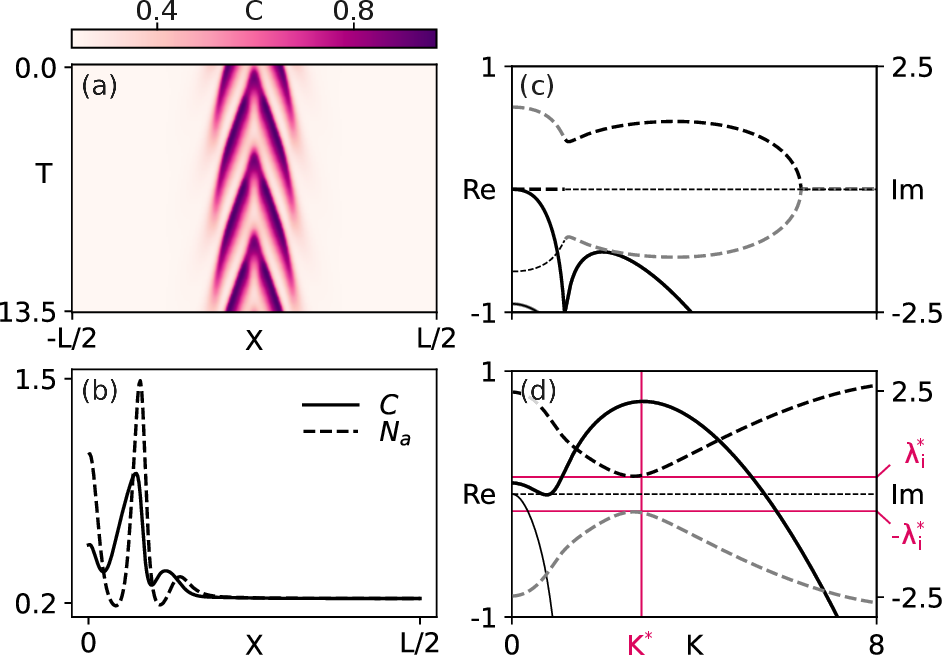}
		\caption{\label{fig:fig_2} Localized oscillatory state at low contractility. (a) Kymograph of the active fluid density $C$. (b) Right half profile of the active fluid and active nucleator densities, at $T=0$ of panel (a). (c, d) Dispersion relations (full lines: real parts, dashed lines: imaginary parts) of Eqs.~\eqref{eq:actin_dynamics}--\eqref{eq:stress_field} linearized around H as a function of the non-dimensional wavenumber $K$. Panels (c) and (d) have the same horizontal axis. Parameters as in Fig.~\ref{fig:fig_1}, with $\Omega_d/\bar{N} = 10$, $\Omega/\bar{N}^2 = 15$, $Z/\bar{N}^2 = B/\bar{N}^3 = 6$, (a--c) $\bar{N} = 1$ and (d) $\bar{N} = \bar{N}_\mathrm{loc} = 1.38$.}
	\end{figure}

	In the vicinity of the critical parameter values for which the homogeneous state H loses linear stability, we numerically find also an instability with respect to localized perturbations of finite amplitude. For low contractility, the states emerging from a localized perturbation are dominantly LOS. One example is illustrated in Fig.~\ref{fig:fig_2}(a,b). For this state, a peak in the concentration of active nucleators emerges periodically in its center. It induces an increase in the density of the active fluid, then broadens, and splits. The two thus created peaks move at constant speed outward until they vanish at the state's boundaries. For these internal ``pulses'', a high nucleator concentration is present at the leading edge, followed by a region of high active fluid density, Fig.~\ref{fig:fig_2}(b). This configuration is similar to that found in actin polymerization waves~\cite{ecker2021} and suggests a similar mechanism behind the motion of the peaks.

	For the parameter values chosen in Fig.~\ref{fig:fig_2}, the state H is linearly stable, Fig.~\ref{fig:fig_2}(c). When increasing the parameter $\Omega$ further, H undergoes a Type $\mathrm{I_o}$ instability. Yet, the origin of the LOSs can be traced back to the instability of a homogenous state. Indeed, the nondimensional parameters $\Omega$ and $Z$ varied in the stability diagram Fig.~\ref{fig:fig_1} depend both on the total average nucleator density $\bar{N}$ as $\Omega,Z\propto\bar{N}^2$. As a consequence, an instability can be induced by increasing the nucleator concentration. If localization of nucleators enhances the nucleator density sufficiently, one may expect the localized state to be patterned. This line of reasoning is similar to the one introduced in Refs.~\cite{herschkowitz-kaufman1972, herschkowitz-kaufman1975a}, where patterns in a reaction-diffusion system subject to heterogeneous forcing were studied.
	
	To explore this idea further, consider the region of high nucleator density, which in the example of Fig.~\ref{fig:fig_2} extends between the boundaries of the localized oscillatory region $X_l = - 5.09$ and $X_r = -X_l$. Concretely, the boundaries are defined as the points at which $N_i$ (the most diffusive, hence spread-out species) deviates more than $10\%$ from its value in the background, that is at $X = \pm L/2$. This criterion was used in Ref.~\cite{barberi2023} to define the boundaries of LSSs. This is a small region containing the localized oscillations in which the total average density of nucleators, $N_\mathrm{loc} = \frac{1}{X_r-X_l} \int_{X_l}^{X_r} (N_a + N_i) dX$, remains constant, Appendix~\ref{sec:appendix_boundaries}.
	
	We can study the linear stability of the homogenous state that has a total density of nucleators equal to that within the localized state, $\bar{N}_\mathrm{loc} = \int_{X_l}^{X_r} (N_a + N_i) dX / (X_l - X_r)$. The corresponding growth exponents show that the homogenous state is linearly unstable and the eigenvalues with largest real part have a nonvanishing imaginary part, Fig.~\ref{fig:fig_2}(d).
	
	It is apparent from Fig.~\ref{fig:fig_2}(d) that the system is well beyond the instability threshold. This is why one should not be surprised by the fact that the oscillation period deduced from the imaginary part of the fastest growing mode, $2\pi/\lambda_i^* = 13.69$, does not capture well the period of the state in Fig.~\ref{fig:fig_2}(a,b), which is $4.54$.

\subsection{High-contractility localized oscillations}\label{sec:high_contractility_los}

	\begin{figure}
		\includegraphics[width=\columnwidth]{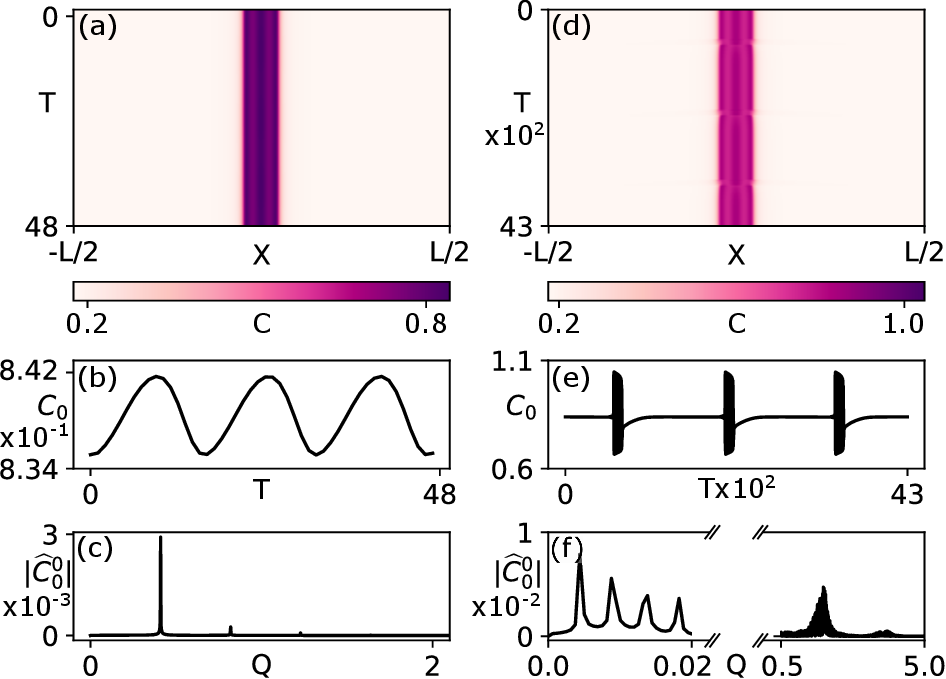}
		\caption{\label{fig:fig_3} Localized oscillatory states at high contractility. (a, d) Kymographs of the active fluid density $C$. (b, e) Time series of the active fluid density $C_0= C(X=0, T)$. (d, f) Fourier spectrum $|\widehat{C_0^0}|$ of $C_0^0 = C_0 - \langle C_0 \rangle_T$, where $\langle C_0 \rangle_T$ is the temporal average of $C_0$. Q is the non-dimensional frequency. The horizontal axis of panel (f) is broken, and two different scales are used on the left and right sides.  Parameters as in Fig.~\ref{fig:fig_1}, with $Z = 15$, $\Omega = 6.665$ (a) and $\Omega = 6.666$ (b).}
	\end{figure}
	
	At higher contractility, LOSs become less frequent, Fig.~\ref{fig:fig_1}, and their profile changes compared to Fig.~\ref{fig:fig_2}(a). Two examples are illustrated in Fig.~\ref{fig:fig_3}. Compared to their low-contractility counterparts, these LOSs are more confined in space. In terms of their time dependence, different states can be distinguished. The state in Fig.~\ref{fig:fig_3}(a--c) exhibits weak oscillations composed of a single frequency and its harmonics. We  refer to these as unimodal states. Conversely, the dependence on time of the state shown in Fig.~\ref{fig:fig_3}(d--f) exhibits broader peaks as well as a band of modes with high frequency. This spectrum corresponds to intermittent bursts on top of regular oscillations. We  refer to these as intermittent states.

\begin{figure}
		\includegraphics[width=\columnwidth]{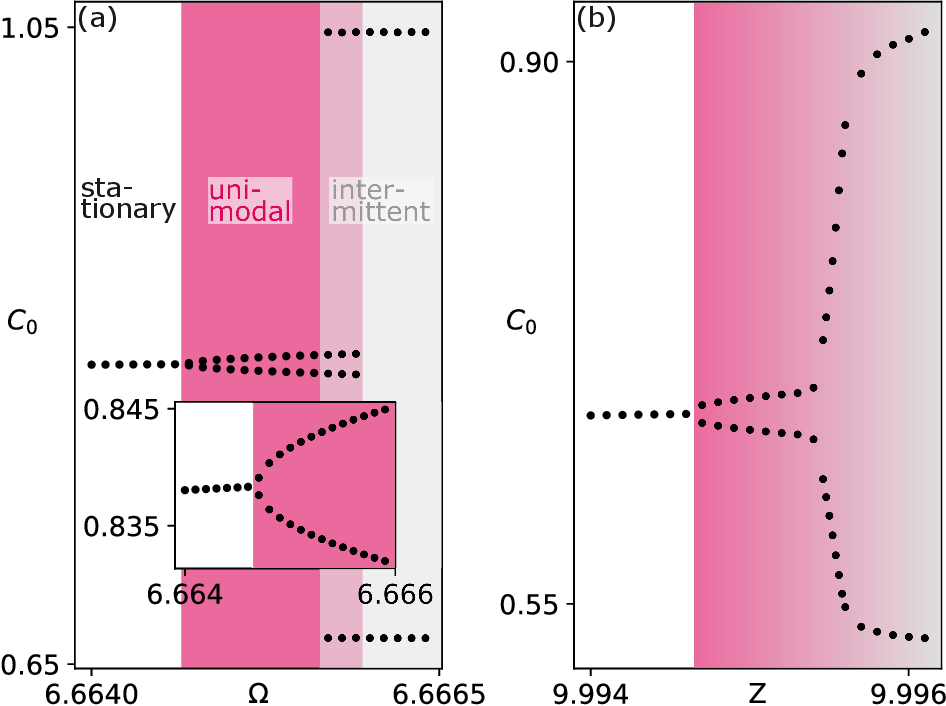}
		\caption{\label{fig:fig_4} Bifurcation diagrams of oscillatory localized states. The two branches of oscillatory states represent, respectively, the maximum and minimum values attained by $C_0$, defined as in Fig.~\ref{fig:fig_3}, during an oscillation. Parameters as in Fig.~\ref{fig:fig_1}, with $Z = 15$ (a) and $\Omega = 9$ (b).}
	\end{figure}

	Differently from the low-contractility case, the LOSs in Fig.~\ref{fig:fig_3} are not the consequence of a local instability of H. Rather, they result from a secondary instability of an LSS. Some insight on this aspect can be gained from the bifurcation diagram in Fig.~\ref{fig:fig_4}(a), which was obtained from numerical integration of the dynamic Eqs.~\eqref{eq:actin_dynamics}--\eqref{eq:stress_field}. Branches were followed by taking the asymptotic solution for some parameter values as initial conditions for slightly different parameter values. Starting from an LSS solution and gradually increasing $\Omega$, a branch of unimodal LOSs emerges supercritically.
	 
	When increasing $\Omega$ further, a branch of intermittent LOSs emerges subcritically. A similar sequence of states can be observed along different directions in parameter space, for instance using $Z$ as control parameter, Fig.~\ref{fig:fig_4}(b). Although the qualitative behavior of the system is similar along the $\Omega$ and the $Z$ direction, the nature of the bifurcations can change. In particular, the transition from unimodal to intermittent LOSs is reminiscent of a canard explosion~\cite{benoit1981}, Fig.~\ref{fig:fig_4}(b).
	
	Further insight into the transition from LSS to LOS can be gained from a linear stability analysis of the LSS. We consider the fields
	\begin{align}
		C &= C^0(X) + \delta C(X,T), \label{eq:pert_C}\\
		N_a &= N_a^0(X) + \delta N_a(X,T), \label{eq:pert_Na}\\
		N_i &= N_i^0(X) + \delta N_i(X,T), \label{eq:pert_Ni}\\
		V &= V^0(X) + \delta V(X,T) \label{eq:pert_V},
	\end{align}
	where the superscript $0$ denotes the LSS solution and the perturbative terms indicated by $\delta$ are assumed to be small compared to the LSS, but of the same order among each other.
	
	We now linearize Eqs.~\eqref{eq:actin_dynamics}--\eqref{eq:force_balance}, by using the expressions in Eqs.~\eqref{eq:pert_C}--\eqref{eq:pert_V} and by only retaining terms linear in the perturbations. We start from the force balance equation, Eq.~\eqref{eq:force_balance}. We get
	\begin{align}
		\partial_X^2 \delta V + \Pi'_0 \partial_X \delta C + \Pi''_0\left( \partial_X C^0\right) \delta C = \delta V, \label{eq:linearized_force_balance}
	\end{align}
	where $\Pi'_0 = \left(\partial \Pi/\partial C \right)\rvert_{C^0}$ and $\Pi''_0 = \left(\partial^2 \Pi/\partial C^2 \right)\rvert_{C^0}$. Equation~\eqref{eq:linearized_force_balance} can be solved for $\delta V$, 
	\begin{align}
		\delta V = (1 - \partial_X^2)^{-1} \left[\left(\Pi'_0 \partial_X + \Pi''_0 \partial_X C^0 \right)\delta C\right], \label{eq:deltaV}
	\end{align}
	where $(1 - \partial_X^2)^{-1}$ represents a formal inversion of the differential operator in  parenthesis. 
	
	We proceed to linearize Eqs.~\eqref{eq:actin_dynamics}--\eqref{eq:nucleator_dynamics_b}. Again, by using the expressions in Eqs.~\eqref{eq:pert_C}--\eqref{eq:pert_V} and retaining linear terms, we get
	\begin{widetext}
		\begin{align}
			\partial_T \delta C + \partial_X \left(C^0 \delta V + V^0 \delta C\right) - \mathscr{D}_c \partial_X^2 \delta C &= \frac{\partial R_c}{\partial C}\Bigg|_{0} + \frac{\partial R_c}{\partial N_a}\Bigg|_{0}, \label{eq:linearized_actin_dynamics} \\
			\partial_T \delta N_a + \partial_X \left(N_a^0 \delta V + V^0 \delta N_a\right) - \mathscr{D}_a \partial_X^2 \delta N_a &= \frac{\partial R_a}{\partial C}\Bigg|_{0} + \frac{\partial R_a}{\partial N_a}\Bigg|_{0} + \frac{\partial R_a}{\partial N_i}\Bigg|_{0},  \label{eq:linearized_nucleator_dynamics_a}\\
			\partial_T \delta N_i + \partial_X \left(N_i^0 \delta V + V^0 \delta N_i\right) - \mathscr{D}_i \partial_X^2 \delta N_i &= \frac{\partial R_i}{\partial C}\Bigg|_{0} + \frac{\partial R_i}{\partial N_a}\Bigg|_{0} + \frac{\partial R_i}{\partial N_i}\Bigg|_{0}, \label{eq:linearized_nucleator_dynamics_b}
		\end{align}
	\end{widetext}
where $R_c = -K_d C + A N_a$, $R_a = - R_i = -\Omega_d C N_a + \Omega_0\left(1 + \Omega N_a^2\right)N_i$, and $\lvert_0$ denotes derivatives evaluated at $(C^0, N_a^0, N_i^0)$.
	
	We analyze Eqs.~\eqref{eq:deltaV}--\eqref{eq:linearized_nucleator_dynamics_b} numerically, by considering the discretized spatial grid $\bm{X} = (X_1, X_2, \dots X_n)$, with spacing $\Delta X$ and $n = 512$ sites. We map spatial fields into vectors with $n$ elements and spatial derivatives into $n \times n$, five-point stencil finite difference operators~\cite{leveque2007a}, $\partial_X^m \to \mathds{D}_m$, where $m$ is the order of the derivative. In Appendix~\ref{sec:appendix_matrix}, we show that Eqs.~\eqref{eq:deltaV}--\eqref{eq:linearized_nucleator_dynamics_b} can thus be mapped onto a $3n \times 3n$ system of ordinary differential equations,
	\begin{align}
		\partial_T 
		\begin{pmatrix}
			\delta \bm{C} \\ 
			\delta \bm{N}_a \\
			\delta \bm{N}_i
		\end{pmatrix}
		=
		\mathds{M}
		\begin{pmatrix}
			\delta \bm{C} \\ 
			\delta \bm{N}_a \\
			\delta \bm{N}_i
		\end{pmatrix}
		= 
		\begin{pmatrix}
			\mathds{M}_{cc} & \mathds{M}_{ca} & \mathds{M}_{ci} \\ 
			\mathds{M}_{ac} & \mathds{M}_{aa} & \mathds{M}_{ai} \\
			\mathds{M}_{ic} & \mathds{M}_{ia} & \mathds{M}_{ii}
		\end{pmatrix}
		\begin{pmatrix}
			\delta \bm{C} \\ 
			\delta \bm{N}_a \\
			\delta \bm{N}_i
		\end{pmatrix},
		\label{eq:matrix}
	\end{align}
	where the vectors are defined as $(\delta C_1, \cdots, \delta C_n, \delta N_{a,1}, \cdots \delta N_{a,n}, \delta N_{i,1}, \cdots \delta N_{i,n})$ and $\mathds{M}_{ij}$, with $i,j \in \{c, a, i\}$, are $n \times n$ blocks.
	
\begin{figure}
		\includegraphics[width=\columnwidth]{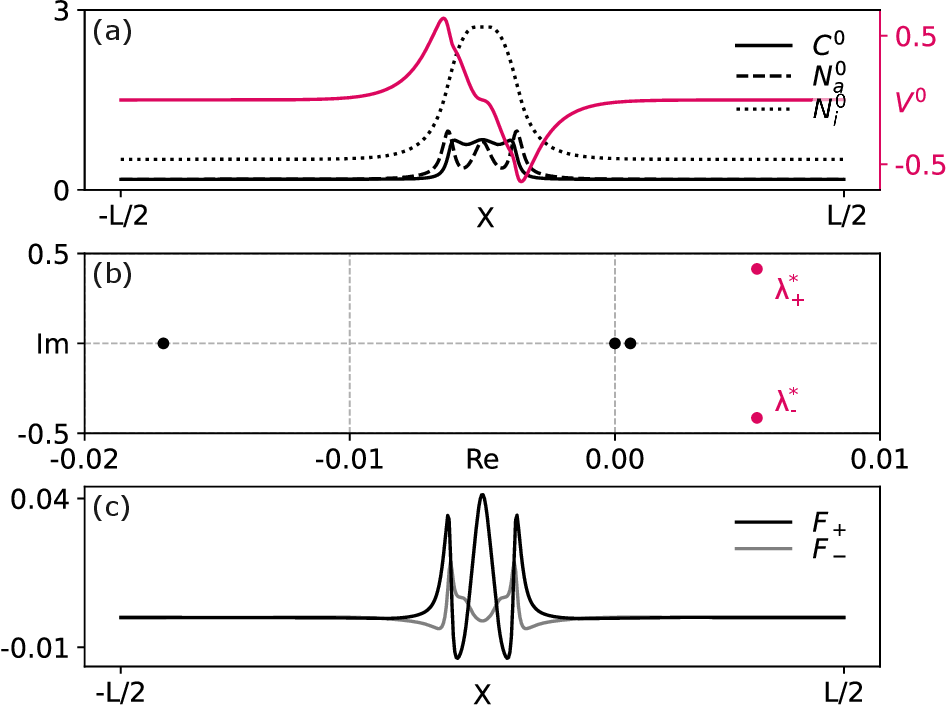}
		\caption{\label{fig:fig_5} Linear oscillatory instability of a localized state. (a) Profile of the LSS for parameter values as in Fig.~\ref{fig:fig_4}(a) and $\Omega=6.664$. (b) Leading eigenvalues of $\mathds{M}$, obtained by linearizing around the LSS in panel (a), for $\Omega = 6.67$. (c) Real-valued $F_+ = \left(f_+ + f_-\right) / 2$ and $F_- = \left(f_+ - f_-\right) / 2i$, where $f_+$ and $f_-$ are the eigenfunctions corresponding to $\lambda_+^*$ and $\lambda_-^*$ in panel (b), respectively. Only the components corresponding to $\delta \bm{C}$ are shown.}
	\end{figure}

	We compute the spectrum of $\mathds{M}$ for the LSS found for $Z=15$ and $\Omega = 6.664$, the profile of which is given in Fig.~\ref{fig:fig_5}(a). At $\Omega = 6.67$, the LSS is linearly unstable. The instability is governed by a pair of complex-conjugate eigenvalues, $\lambda_{\pm}^*= \lambda_r^* \pm i\lambda_i^*$, with $\lambda_r^* > 0$, Fig.~\ref{fig:fig_5}(b). Note that the eigenvalue $\lambda=0$ results from the conservation of nucleators. We have verified that these results are qualitatively robust to changes in $n$. However, we note that the eigenvalues can be very sensitive to $\Omega$. Changing $n$ when $\Omega$ is close to its critical value can qualitatively change the stability of the LSS.
	
	At $\Omega = 6.67$, we estimate the oscillation period $\tau$ from the imaginary part of the eigenvalues,  $\tau\approx2\pi / \lambda^*_{\pm, i} = 15.17$. This value is in very good agreement with the dominant mode of the LOS in Fig.~\ref{fig:fig_3}(a--c), for which $\tau = 15.34$. To further connect the LOS with the LSS, we consider the eigenfunctions $f_+$ and $f_-$ corresponding to $\lambda^*_\pm$. Through suitable linear combinations of $f_+$ and $f_-$, we obtain real-valued functions $F_+ = \left(f_+ + f_-\right) / 2$ and $F_- = \left(f_+ - f_-\right) / 2i$. Their components corresponding to the density perturbation $\delta\bm{C}$ are given in Fig.~\ref{fig:fig_5}(c). These functions are spatially localized within a region of width comparable to that of the LSS in Fig.~\ref{fig:fig_5}(a). Close to the instability, linear stability analysis provides an approximate expression of the oscillatory state, Appendix~\ref{sec:appendix_eigenfunctions}. This confirms that the LOS in Fig.~\ref{fig:fig_3}(a--c) emerges from a localized instability.
	
\subsection{Localized chaotic-like dynamics}

\begin{figure}
	\includegraphics[width=\columnwidth]{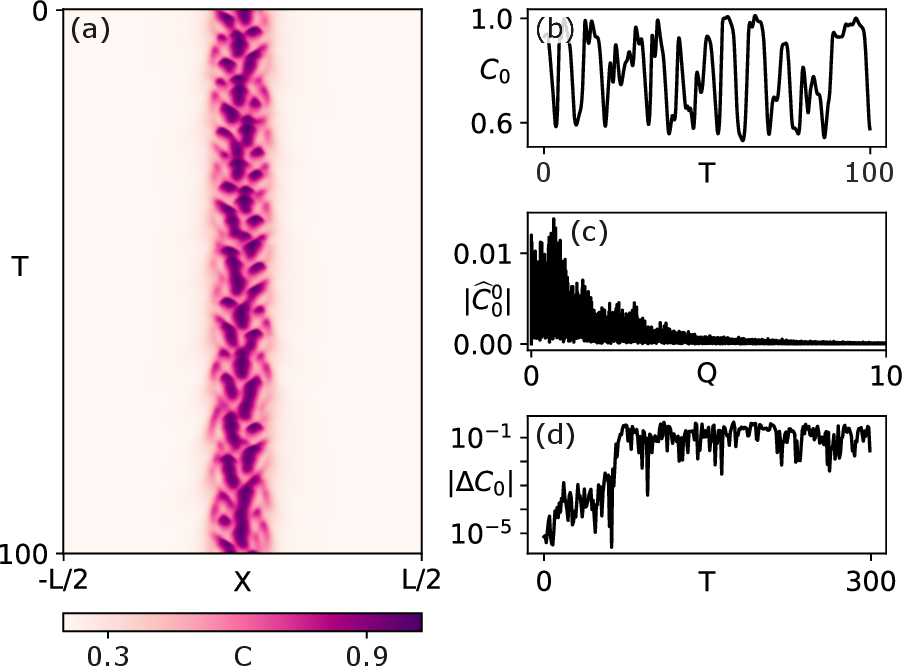}
	\caption{\label{fig:fig_6} Localized chaotic-like state.  (a) Kymograph of the active fluid density $C$. (b) Time series of the active density $C_0$, defined as in Fig.~\ref{fig:fig_3}. (c) Fourier spectrum as defined in Fig.~\ref{fig:fig_3}. (d) Absolute value of the difference between two time series $C_0$ with similar initial conditions. Vertical axis is in log scale. Parameters as in Fig.~\ref{fig:fig_1}, with $Z = 9.3$ and $\Omega = 12$.}
\end{figure}

In addition to the periodic LOSs discussed thus far, the dynamic Eqs.~\eqref{eq:actin_dynamics}--\eqref{eq:nucleator_dynamics_b} also have more complicated localized solutions, Fig.~\ref{fig:fig_6}(a). The width of the state shown is similar to that of LSS or LOS for similar values of $Z$. However, the concentration profile of the active fluid is irregular. The time-dependence of the active-fluid density $C_0 = C(X = 0, T)$ does not exhibit an easily recognizable pattern, Fig.~\ref{fig:fig_6}(b), which is also expressed by its broad Fourier spectrum, Fig.~\ref{fig:fig_6}(c).  
	
This irregular localized state is likely chaotic. Indeed, for the corresponding parameter values, the absolute difference between $C_0$ for two slightly different initial conditions initially grows exponentially, Fig.~\ref{fig:fig_6}(d). Specifically, we took the initial condition of the state shown in Fig.~\ref{fig:fig_6}(a) and added two different weak white noise perturbations to it. Eventually, this difference saturates, which is an effect of the finite size of the attractor~\cite{strogatz2019}. 

An in-depth characterization of this state, which notably addresses the possibility that it is transient, as well as an exploration of the route to chaos is left for further studies.
	
\section{Localized spatiotemporal dynamics in two dimensions}
\label{sec:twoD}

\begin{figure}
	\includegraphics[width=0.75\columnwidth]{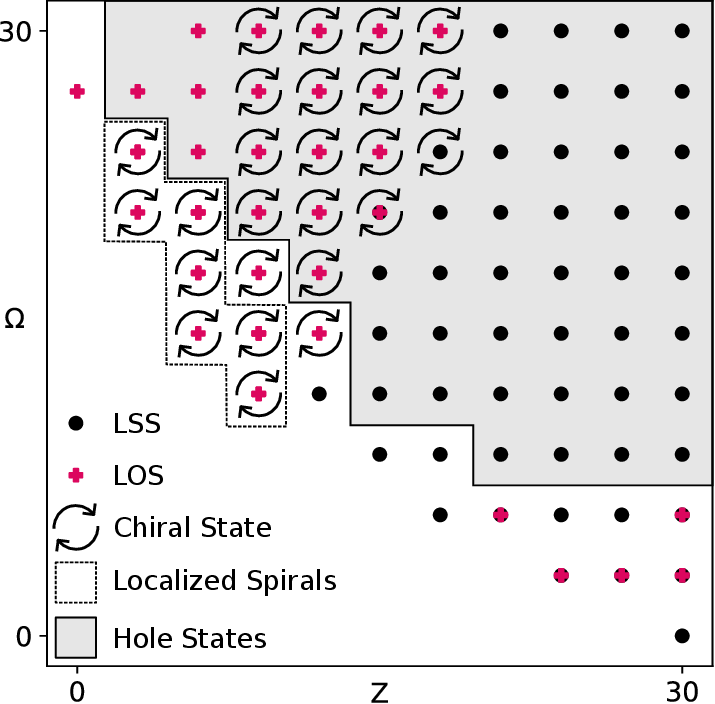}
	\caption{\label{fig:fig_7a} Phase diagram of localized states in two dimensions. Symbols indicate parameter values for which localized states are numerically obtained through time evolution of localized initial conditions. Localized states of different nature and symmetry can coexist. The chiral state symbol denotes the existence of at least one type of chiral state. Localized spirals exist within the region enclosed by the dotted line. Hole states exist within the gray region. Parameters as in Fig.~\ref{fig:fig_1}. Numerical solutions are obtained on a grid of $512 \times 512$ sites.}
\end{figure}

Similar to LSSs \cite{barberi2023}, LOSs persist in two spatial dimensions (2D). Much of the discussion of LOSs in 1D readily extends to the 2D case. Consider a square domain, $(X, Y) \in [-L/2, L/2] \times [-L/2, L/2]$, with periodic boundaries and $L=10\pi$. The distribution of some 2D LSSs and LOSs in a region of parameter space is illustrated in Fig.~\ref{fig:fig_7a}. 

\begin{figure}
	\includegraphics[width=0.85\columnwidth, height = 0.7\columnwidth]{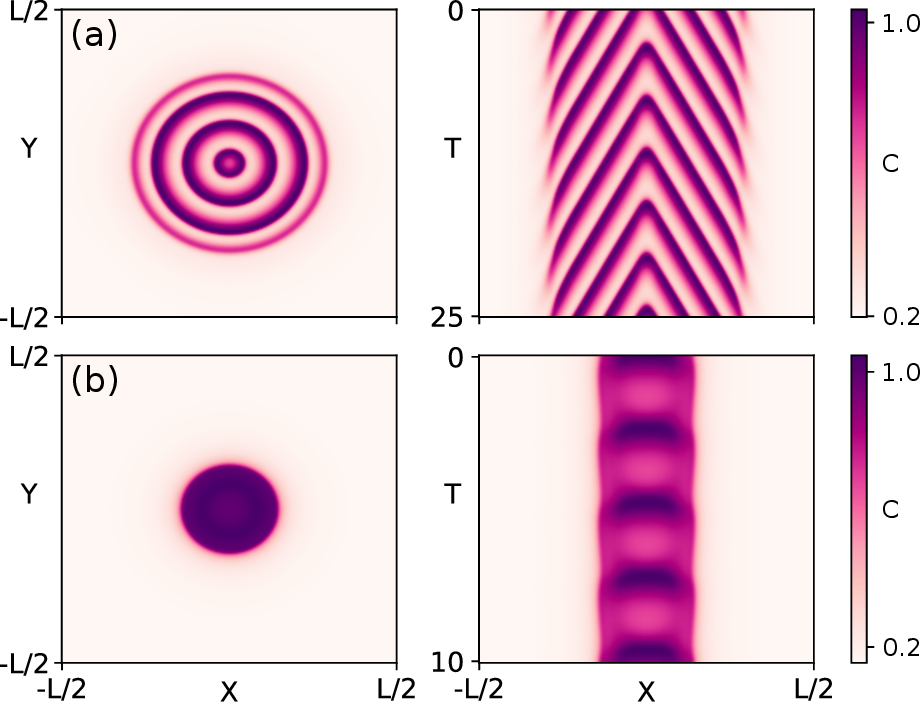}
	\caption{\label{fig:fig_7} Isotropic localized oscillatory states in two dimensions. (a) Low-contractility, localized target wave. (b) High-contractility, pulsating spot. (a--b) The kymographs on the right represent a cut of the 2D dynamics at $Y = 0$. All panels have the same horizontal axis. Parameters as in Fig.~\ref{fig:fig_1}, with $Z = 9$, $\Omega = 15$ (a), $Z = 24$, $\Omega = 3$ (b).}
\end{figure}

For low contractility, LOSs can take the form of target waves, Fig.~\ref{fig:fig_7}(a). Note the similarity between the kymograph in Fig.~\ref{fig:fig_7}(a) and that in Fig.~\ref{fig:fig_2}(a). Similarly to the 1D case, the maxima in active fluid density move outwards at constant velocity. The oscillation frequency is similar in 1D and 2D, whereas the spatial extension of the 2D state shown in Fig.~\ref{fig:fig_7}(a) is larger than for the 1D state of Fig.~\ref{fig:fig_2} in spite of a larger contractility, $Z=9$ in 2D \textit{vs} $Z=6$ in 1D. For higher contractility, LOSs can take the form of pulsating spots, Fig.~\ref{fig:fig_7}(b). Similarly to the 1D case, these feature a stronger confinement as compared to their low-contractility counterparts.

\begin{figure}
	\includegraphics[width=0.85\columnwidth, height = 0.7\columnwidth]{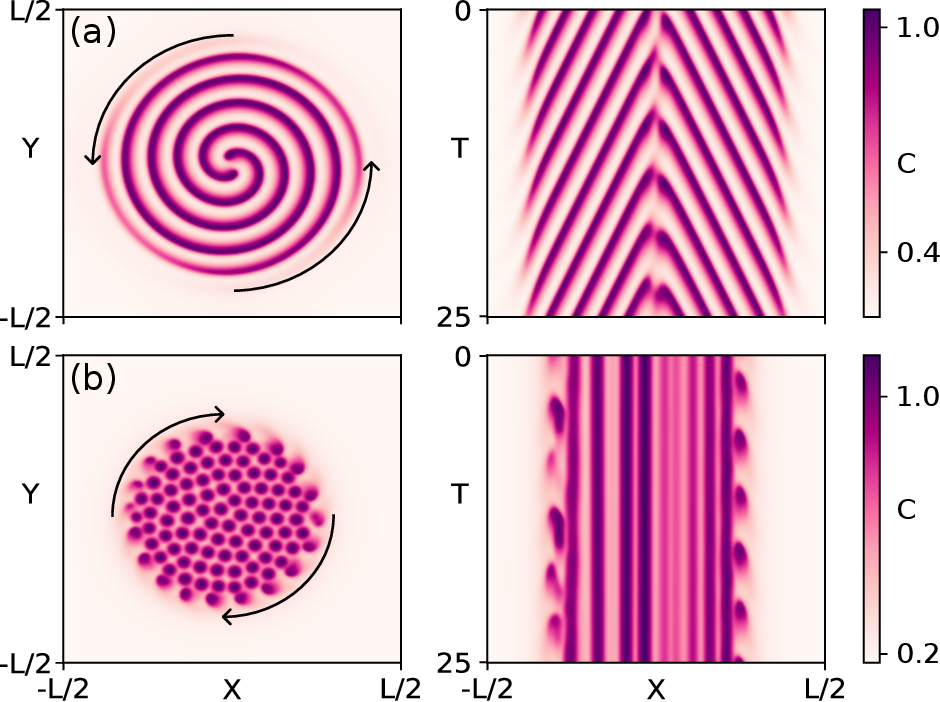}
	\caption{\label{fig:fig_2d_chiral} Localized chiral states in two dimensions. (a) Localized double-arm spiral rotating counterclockwise (see arrows). (b) Closely packed spots, surrounded by a crown of sparser spots rotating clockwise (see arrows). (a--b) The kymographs on the right represent a cut of the 2D dynamics at $Y = 0$. All panels have the same horizontal axis. Parameters as in Fig.~\ref{fig:fig_1}, with $\Omega = 21$ and $Z = 6$ (a), $\Omega = 15$ and $Z = 12$ (b).}
\end{figure}

\begin{figure}
	\includegraphics[width=0.85\columnwidth]{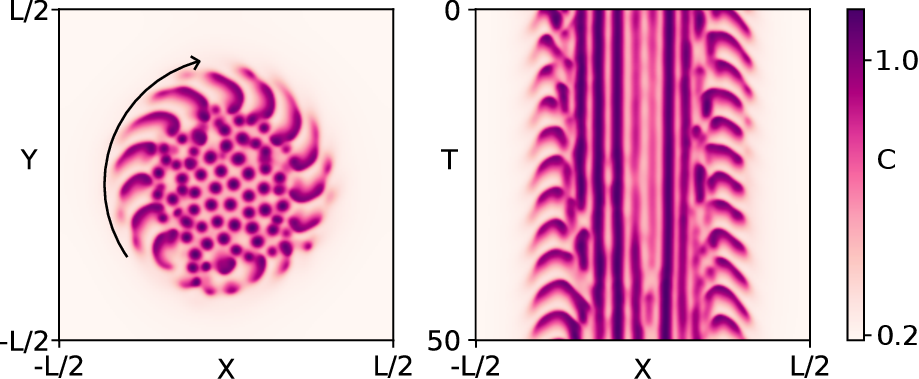}
	\caption{\label{fig:fig_2d_chaotic} Dynamic multi-spot core surrounded by waves. The arrow indicates the direction of some wavefronts. The kymograph on the right represents a cut of the 2D dynamics at $Y = 0$. All panels have the same horizontal axis. Parameters as in Fig.~\ref{fig:fig_1}, with $Z = 9$ and $\Omega = 18$.}
\end{figure}
 
There are also LOSs in 2D that do not have a 1D equivalent. In particular, localized states in 2D can spontaneously break chiral symmetry. At relatively low contractility and chemical activity, this generates localized spirals, Fig.~\ref{fig:fig_7a} and Fig.~\ref{fig:fig_2d_chiral}(a). As either contractility or chemical activity increases, localized spirals develop a core of closely packed, almost stationary spots, surrounded by a crown of spots traveling around the central cluster, Fig.~\ref{fig:fig_2d_chiral}(b). These traveling spots are less dense than the spots in the cluster, and feature a higher active fluid density towards their direction of motion. 2D LOSs with a broader traveling crown and a less regular core are also possible, Fig.~\ref{fig:fig_2d_chaotic}.

\begin{figure}
	\includegraphics[height = 0.7\columnwidth]{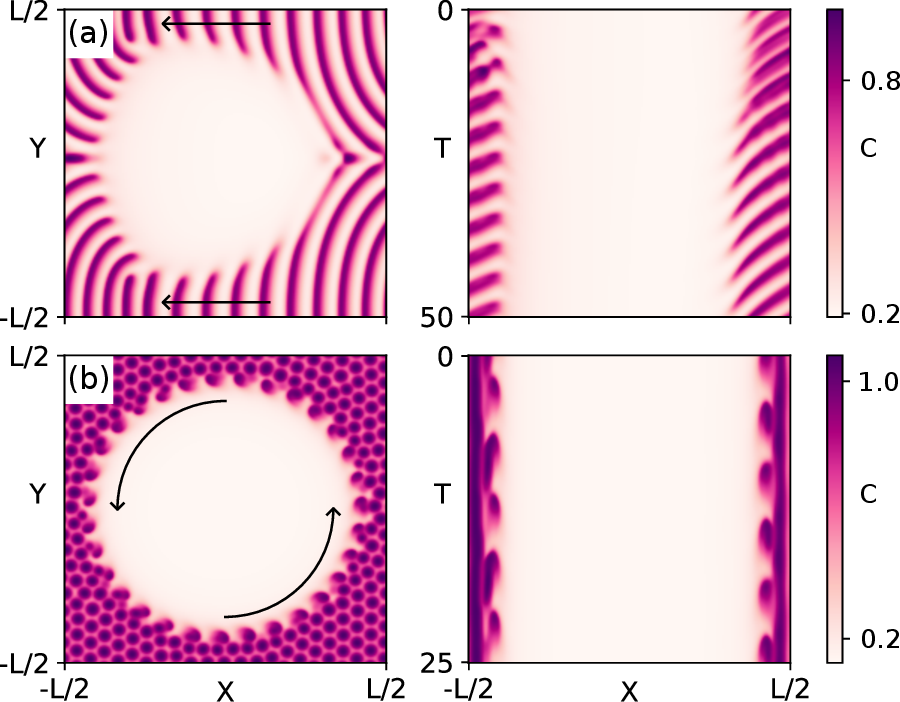}
	\caption{\label{fig:fig_2d_holes} Hole states in two dimensions. (a) Hole in a background of traveling wave fronts. The arrows indicate the local direction of motion of some wave fronts. Over time, the hole drifts leftwards with speed $\simeq 0.01$. (b) Hole in a background of closely packed spots. The interface between the homogeneous state and the closely-packed spots consists of a crown of sparser spots, which rotate counterclockwise (see arrows). (a--b) The kymographs on the right represent a cut of the 2D dynamics at $Y = 0$. All panels have the same horizontal axis. Parameters as in Fig.~\ref{fig:fig_1}, with $\Omega = 24$ and $Z = 6$ (a), $\Omega = 24$ and $Z = 12$ (b).}
\end{figure}

In addition to localized states where a spatiotemporal pattern is surrounded by the homogeneous steady state, a rich variety of ``hole states'' exists in 2D, where the homogeneous steady state is surrounded by a spatiotemporal pattern. Hole states emerge in a large portion of parameter space, Fig.~\ref{fig:fig_7a}, and generally coexist with the localized states described above. Hole states can feature both static and dynamic background states. The background pattern depends on parameter values and, in general, it resembles the localized pattern observed for the same parameter values. For instance, at low contractility, we find holes in a background of traveling fronts, Fig.~\ref{fig:fig_2d_holes}(a). Conversely, at high contractility, we find holes in a background of closely packed spots, Fig.~\ref{fig:fig_2d_holes}(b). Hole states can also break chiral symmetry, Fig.~\ref{fig:fig_2d_holes}(b), resulting in chiral spatiotemporal patterns similar to those described above.
		
\section{Conclusions}

In this work, we have reported LOSs in a mechanochemical system, where active nucleators promote the assembly of an active fluid. Nucleators are advected by fluxes induced by active stress and are inactivated by the active fluid. We identified two possible origins of LOSs, namely, a local instability of the homogenous steady state and a secondary instability of LSSs.

LOSs similar to the ones discussed above were reported in a broad range of dissipative systems. For instance, in vertically vibrated granular layers~\cite{umbanhowar1996}, where they were termed `oscillons'. LSS-to-LOS instabilities and localized chaos were also reported in reaction-diffusion systems~\cite{alsaadi2021} and optical cavities~\cite{leo2013, parra-rivas2021}. Localized spatiotemporal chaos was realized in liquid crystals~\cite{verschueren2013}, where the resulting state was termed `chaoticon'. The chiral state in Fig.~\ref{fig:fig_2d_chiral}(b) is reminiscent of the ratcheting states observed in ``cellular flames''~\cite{markstein1949, gorman1996}.

Our choice of dynamic equations for the mechanochemical system is not unique~\cite{bois2011,kumar2014,banerjee2017,nishikawa2017,staddon2022,deljunco2022}, and it will be interesting to investigate, which of the phenomena we report depend on details of the coupling between biochemistry and active mechanics. In the parameter regions we investigated, localization depends crucially on (active) contractility. We expect that a system exhibiting these two features -- chemically induced oscillations and contractility -- will generically have LOSs as solutions. Still, the explicit dependence of the reaction rates and of the stress on the densities might affect these states.

Our theory is inspired by the actin cortex of animal cells. Following the remarks in the previous paragraph, localized oscillations could be a common feature of cortical actin. The LOSs we found provide a new perspective on cortical structures like transient filamentous protrusions~\cite{cai2017}, or the states reported in Refs.~\cite{graessl2017,baird2017}. Our study thus suggests a strong link between these LOSs and spontaneous actin waves. By carefully tuning cortical contractility, cells might transition between these two classes of states. More experiments are needed to explore this connection further.

	\begin{acknowledgments}
		We thank Damien Brunner, Olivier Pertz, Daniel Riveline and their groups, as well as Alan Champneys, Ludovic Dumoulin, Nicolas Ecker, Oriane Foussadier, Lendert Gelens, and Edgardo Villar Sepulveda for useful discussions. Numeric calculations were performed at the University of Geneva on the “Baobab” HPC cluster. This work was funded by Swiss National Science Foundation Sinergia grant CRSII5\_183550.
	\end{acknowledgments}
	
	\appendix
	
	\section{Boundaries of the high-nucleator-density region}\label{sec:appendix_boundaries}
	
\begin{figure}
		\includegraphics[width=0.75\columnwidth]{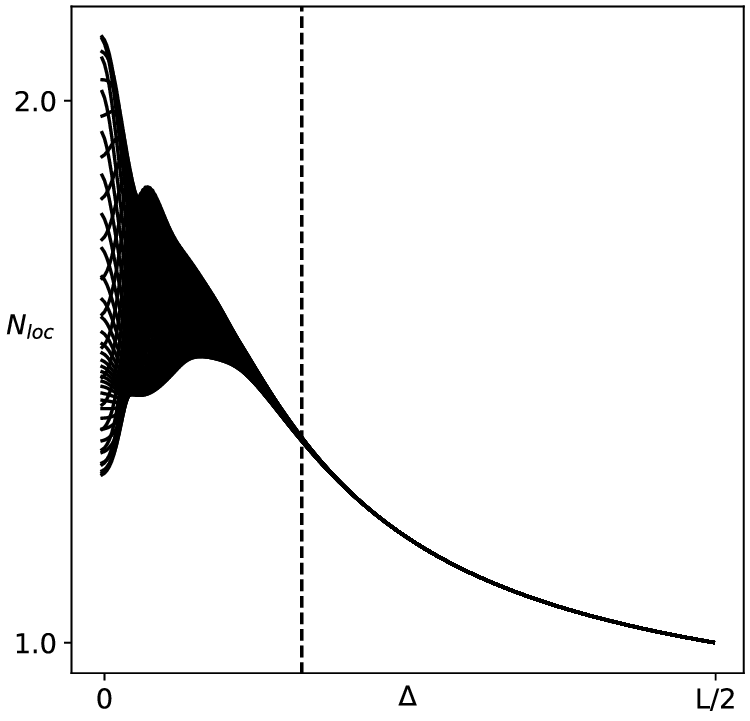}
		\caption{\label{fig:appendix_boundaries} Boundaries of a localized oscillatory state. Definitions and details in Sec.~\ref{sec:low-contractility_los} and Appendix~\ref{sec:appendix_boundaries}.}
	\end{figure}

	To illustrate that the 10\% criterion presented in Sec.~\ref{sec:low-contractility_los} is well suited to studying LOSs, here we study the total average nucleator density, $N_\mathrm{loc}$, in a region around the localized oscillations in Fig.~\ref{fig:fig_2}. We define $X_l=-\Delta$ and $X_r=\Delta$, with $\Delta \in [0, L/2]$, and plot $N_\mathrm{loc} = N_\mathrm{loc}(\Delta)$. In Fig.~\ref{fig:appendix_boundaries}, the different lines represent $N_\mathrm{loc}(\Delta)$ at different times, over an oscillation period. As $\Delta \to L/2$, $N_\mathrm{loc} \to 1$, because the interval $[X_l,X_r]$ becomes as large as the whole system. As $\Delta \to 0$, $N_\mathrm{loc}$ undergoes strong temporal and spatial fluctuations, respectively due to the localized state’s temporal oscillations and spatial patterning. The boundary between the two regimes is captured by the vertical dashed line, obtained via the $10\%$ criterion.
	
	\section{Discretized dynamical equations}\label{sec:appendix_matrix}
	Upon discretization, Eq.~\eqref{eq:deltaV} provides a simple expression of $\delta\bm{V}$ in terms of $\delta\bm{C}$,
	\begin{align}
		\delta \bm{V} &= \mathds{S} \delta \bm{C}, \label{eq:pert_V_2} \\
		\mathds{S} &= \left(\mathds{1} - \mathds{D}_2\right)^{-1} \left[\mathrm{di}\left(\bm{\Pi}'_0\right)\mathds{D}_1 +  \mathrm{di}\left(\bm{\Pi}''_0\right) \mathrm{di}\left(\mathds{D}_1 \bm{C}_0\right) \right]. \label{eq:S}
	\end{align}
	Here, $\mathrm{di}(\bm{a})$ is an $n \times n$ matrix that is defined for a vector $\bm{a}$ as $\mathrm{di}(\bm{a})_{ij} = a_i$ if $i=j$ and $= 0$ if $i \neq j$. Also, $\bm{\Pi}'_0$ and $\bm{\Pi}''_0$ represent the discretized spatial fields $\Pi'_0$ and $\Pi''_0$, respectively.
	
	We use Eq.~\eqref{eq:pert_V_2} to eliminate the velocity field from Eqs.~\ref{eq:linearized_actin_dynamics}--\eqref{eq:linearized_nucleator_dynamics_b}, getting the discretized linear system in Eqs.~\ref{eq:matrix}. The entries of the matrix $\mathds{M}$ are:
		\begin{align}
			\mathds{M}_{cc} =& - \di{\mathds{D}_1\bm{C}^0}\mathds{S} -\di{\bm{C}^0}\mathds{D}_1\mathds{S} - \di{\mathds{D}_1 \bm{V}^0}\nonumber\\
			& -\di{\bm{V}^0}\mathds{D}_1 + \mathscr{D}_c\mathds{D}_2 + \di{\frac{\partial R_c}{\partial C}\Bigg|_{0}} \\
			\mathds{M}_{ca} = &~ \di{\frac{\partial R_c}{\partial N_a}\Bigg|_{0}} \\
			\mathds{M}_{ci} = &~ 0 \\
			\mathds{M}_{ac} = & - \di{\mathds{D}_1\bm{N}^0_a}\mathds{S} -\di{\bm{N}_a^0}\mathds{D}_1\mathds{S} + \di{\frac{\partial R_a}{\partial C}\Bigg|_{0}} \\
			\mathds{M}_{aa} = & - \di{\mathds{D}_1 \bm{V}^0} -\di{\bm{V}^0}\mathds{D}_1 + \mathscr{D}_a \mathds{D}_2 \nonumber\\
			&+ \di{\frac{\partial R_a}{\partial N_a}\Bigg|_{0}} \\
			\mathds{M}_{ai} = &~ \di{\frac{\partial R_a}{\partial N_i}\Bigg|_{0}} \\
			\mathds{M}_{ic} = & - \di{\mathds{D}_1\bm{N}^0_i}\mathds{S} -\di{\bm{N}_i^0}\mathds{D}_1\mathds{S} + \di{\frac{\partial R_i}{\partial C}\Bigg|_{0}} \\
			\mathds{M}_{ia} = &~ \di{\frac{\partial R_i}{\partial N_a}\Bigg|_{0}} \\
			\mathds{M}_{ii} = & - \di{\mathds{D}_1 \bm{V}^0} -\di{\bm{V}^0}\mathds{D}_1 + \mathscr{D}_i \mathds{D}_2 + \di{\frac{\partial R_i}{\partial N_i}\Bigg|_{0}}
		\end{align}
	When the operator $\di{\cdot}$ is applied to a function of $X$, we implicitly assume a spatial discretization of that function on the grid $\bm{X}$.
	
	\section{Approximate expression of the oscillatory state close to the instability.}\label{sec:appendix_eigenfunctions}
	For simplicity, we define $\bm{u} = (\delta\bm{C}, \delta\bm{N}_a, \delta\bm{N}_i)$, such that Eq.~\eqref{eq:matrix} reads
	\begin{align}
		\partial_T\bm{u} = \mathds{M}\bm{u}. \label{eq:appendix_eigenfunctions_1}
	\end{align}
	We denote by $\{\bm{f}_i\}_{i=1 \dots 3n}$ the set of eigenvectors of $\mathds{M}$, and by $\{\lambda_i\}_{i=1 \dots 3n}$ their corresponding eigenvalues. Note that the eigenvector $\bm{f}_i$ corresponds to the spatially discretized eigenfunction $f_i$ in Sec.~\ref{sec:high_contractility_los}.
	
	If the eigenvectors form a complete basis, we can express $\bm{u}$ as $\bm{u} = \sum_{i = 1}^{3n} c_i \bm{f}_i$. By plugging this expression into Eq.~\eqref{eq:appendix_eigenfunctions_1} and by assuming that the eigenvectors are orthogonal, we get
	\begin{align}
		\bm{u} = \sum_{i = 1}^{3n} A_i e^{\lambda_i T}\bm{f}_i, \label{eq:appendix_eigenfunctions_2}
	\end{align}
	where $A_i$ are constants.
	
	As $T\to\infty$, the sum in Eq.~\eqref{eq:appendix_eigenfunctions_2} is dominated by the terms corresponding to the eigenvalue with the largest real part. In our case, there is not a single eigenvalue, but rather a complex-conjugate pair, $\lambda^*_\pm$, with corresponding eigenvectors $\bm{f}_\pm$. Therefore, asymptotically,
	 \begin{align}
	 	\bm{u} \simeq A_+ e^{\lambda_+ T}\bm{f}_+ + A_- e^{\lambda_- T}\bm{f}_-. \label{eq:appendix_eigenfunctions_3}
	 \end{align}
	 
	 Note that, since $\mathds{M}$ is a real matrix, $\bm{f}_+$ and $\bm{f}_-$ are complex-conjugates. Then, since $\bm{u} \in \mathds{R}^n$, Eq.~\eqref{eq:appendix_eigenfunctions_3} at $T=0$ implies that $A_+ = A_-$. Hence,
	 \begin{align}
	 	\bm{u} \propto e^{\lambda_+ T}\bm{f}_+ + e^{\lambda_- T}\bm{f}_-. \label{eq:appendix_eigenfunctions_4}
	 \end{align}
	 
	 Simple algebra leads from Eq.~\eqref{eq:appendix_eigenfunctions_4} to
	 \begin{align}
	 	\bm{u} \propto e^{\lambda^*_r T}\left[\cos\left(2\pi T/\tau\right)\bm{F}_+ + \sin\left(2\pi T/\tau\right)\bm{F}_-\right], \label{eq:appendix_eigenfunctions_5}
	 \end{align}
	 where $\bm{F}_\pm$ is the spatially discretized eigenfunction $F_\pm$ (defined in Sec.~\ref{sec:high_contractility_los}, together with $\tau$). 
	 
	 In conclusion, an approximate expression for the localized oscillatory state is given by
	 \begin{align}
	 	(\bm{C}^0, \bm{N}_a^0, \bm{N}_i^0) + \bm{u}. 
	 \end{align}
	 The exponential growth of $\bm{u}$ persists only for a finite time after which its amplitude saturates.	 
	
	\bibliography{bibliography}
	
\end{document}